\documentclass [12pt]{article}
\usepackage {amssymb}
\usepackage {amsmath}
\usepackage {graphicx}
\usepackage {longtable}

\begin{document}
\noindent{\footnotesize{{\it International Journal 
of Anticipatory Computing Systems} {\bf 16}, pp. 3-24 (2004) \\
(Also http://arXiv.org/abs/physics/0309102)\\
Ed. D. Dubois, Publ. by CHAOS, Volume 16, 2004, pp. 3-24. ISSN
1373-5411 ISBN 2-930396-02-4. Delivered at CASYS'2003 
(The Conference on Anticipatory Computing Systems, 
Institute of Mathematics, Lieg\`e, Belgium, 11-16 August 2003).}}

\bigskip
\begin{center}
\section*{The Universe from Nothing: \\
A Mathematical Lattice of Empty Sets}
\end{center}

\medskip

\begin{center}
{\fbox{\bf Michel BOUNIAS$^1$}}
\end{center}
\begin{center}
{\bf{and Volodymyr KRASNOHOLOVETS$^2$}}
\end{center}

\begin{center}
$^{1}$\textit{BioMathematics Unit, d'Avignon University and INRA,
France}  \\
$^{2}$\textit{Institute for Basic Research, 90 East Winds Court,
 Palm Harbor, FL 34683, USA}
\end{center}

\medskip
\bigskip
\noindent \textbf{Abstract}

\noindent Major principles of mathematical constitution of space
and the principles of construction of physical space are
presented. The existence of a Boolean lattice with fractal
properties originating from nonwellfounded properties of the empty
set is demonstrated. Space-time emerges as an ordered sequence of
mappings of closed 3-D Poincar${\rm \acute{e}}$ sections of a
topological 4-space-time provided by the lattice of primary empty
cells. The fractal kernel stands for a particle and the reduction
of its volume is compensated by morphic changes of a finite number
of surrounding cells. Quanta of distances and quanta of fractality
are demonstrated. Deformation attributes associated to mass
determine the inert mass and the gravitational effects, but
fractal deformations of cells are responsible for such
characteristics as spin and charge.

\medskip

\noindent \textbf{Key words:} space structure, topological
distances, space-time differential element, fractality, particles

\medskip

\subsection*{1 Introduction}

\hspace*{\parindent} In the area of sub atomic physics
experimental sciences can only outline general peculiarities of
hidden features of the observable world. That is why basic
difficulties in the construction of a unified pattern of the
physical world fall on theoreticians' shoulders. A great number of
approaches to the description of the microworld have been proposed
and a majority of the approaches have established a new branch in
mathematical physics, so-called ``physical mathematics''.
Nevertheless, despite the striking progress in present-day
research, from particle physics to cosmology, many fundamental
questions remain unsolved and often contradictory (Krasnoholovets,
2003a). A principal characteristic feature of modern theories
remains an exceptional abstraction: They are still based on
notions, which themselves must be clarified. For instance: What is
the wave $\psi $-function in quantum mechanics, what is the nature
of quantum fields used in quantum field theories, how can one
understand 10 dimensions in string theories, are there Higgs
bosons in the real physical space or they exist only in abstract
physical constructions, what is the electric charge?, etc.

Contrary to physical mathematics, Bounias (2000a), starting from pure
mathematics, namely, the mathematical theory of demonstration, showed that
any property of any given object should be consistent with the
characteristics of the corresponding embedding space. In other words, the
study of any object requires a preliminary knowledge of its properties, or
its structure. M. Bounias (Bounias and Krasnoholovets, 2003a, b, c) carried
out a tremendous work reconsidering fundamental concepts regarding such
fundamental notions as measure, distance, dimension and fractality, which
were developing in classical mathematics within the whole 20$^{th}$ century.
Bounias (2001) showed that using the biological brain's system, due to its
property of self-decided anticipatory mental imaging, one could overpass
mathematical limits in computed systems. It will be shown below how this
makes eventually possible a scanning of an unknown universe by a part of
itself represented by an internal observer.

Our approach is aimed at searching for distances that would be
compatible with both the involved topologies and the scanning of
objects not yet known in the studied spaces. No such configuration
is believed to be an exception in a general case. Generalized
conceptions of distances and dimensionality evaluation are
introduced, together with their conditions of validity and range
of application to topological spaces. It is argued (Bounias and
Krasnoholovets, 2003a) that the empty set forms a Boolean lattice
with fractal properties and that the lattice provides a substrate
with both discrete and continuous properties. Space-time emerges
as an ordered sequence of mappings of closed 3-D Poincar${\rm
\acute{e}}$ sections of a topological 4-space-time ensured by the
lattice of primary empty cells. A combination rule determining
oriented sequences with continuity of set-distance function
produces a particular kind of 'space-time'-like structure that
favors the aggregation of such deformations into fractal forms
standing for massive objects. The role of the fractality in
deformations of cells, which generates such fundamental
characteristics as charges and electric and magnetic properties,
is investigated in detail.

\subsection*{2. Preliminaries}

\subsubsection*{2.1. Measure, Distances and Dimensions}

\hspace*{\parindent} One of the problems faced by modelling
unknown worlds could be called "the syndrome of polynomial
adjustment". In effect, given an experimental curve representing
the behavior of a system whose real mechanism is unknown, one can
generally perform a statistical adjustment with using a polynomial
system like $\rm M{\kern 1pt} {\kern 1pt} = {\kern 1pt} {\kern
1pt} \sum\nolimits_{( {{\kern 1pt} i{\kern 1pt} = {\kern 1pt}
0{\kern 1pt} {\kern 1pt} \to N{\kern 1pt}}  )} {a_{i}} \cdot
x^{i}.$ Then, using a tool to test for the fitting of the $\rm N +
1$ parameters to observational data will require increasingly
accurate adjustment, so as to convincingly reflect the natural
phenomenon within some boundaries. However, if the real equation
is mathematically incompatible with the polynomial, there will
remain some irreducible parts in the fitting attempts.

\bigskip

\noindent  {2.1.1. Measure}

\bigskip

The concept of measure usually involves such particular features
as the existence of mappings and the indexation of collections of
subsets on natural integers. Classically, a measure is a
comparison of the measured object with some unit taken as a
standard.

The "unit used as a standard", this is the part played by a gauge
(J). Again, a gauge is a function defined on all bounded sets of
the considered space, usually having non-zero real values, such
that (Tricot, 1999): (i) a singleton has measure naught: $\rm
\forall \,x,\;J\left( {\{ x\}}  \right) = 0;$ (ii) (J) is
continued with respect to the Hausdorff distance; (iii) (J) is
growing: $\rm E \subset F \Rightarrow \quad J\left( {E} \right)
\subset  J\left( {F} \right);$ (iv) (J) is linear: $\rm F\left( {r
\cdot E} \right): \ r \cdot J\left( {E} \right).$ This implies
that the concept of distance is defined: usually, a diameter, a
size, or a deviation are currently used, and it should be pointed
that such distances need to be applied on totally ordered sets.
Even the Caratheodory measure $\left( {\mu ^{\ast} } \right)$
poses some conditions that again involve a common gauge to be
used: (i) $\rm A \subset B  \Rightarrow \mu ^{\ast} \left( {A}
\right) \leqslant \mu ^{\ast} \left( {B} \right);$ (ii) for a
sequence of subsets $\rm \left( {Ei} \right): \mu ^{\ast} \cup
\left( {Ri} \right) \leqslant \sum {\mu ^{\ast} \left( {Ri}
\right);} $ (iii) $\rm \angle \left( {A,B} \right),\; A \cap B =
\O {\kern 1pt}$ : $\rm  \mu ^{\ast} \left( {A \cup B} \right) =
\mu ^{\ast} \left( {A} \right) + \mu ^{\ast} \left( {B} \right);$
(iv) $\rm \mu ^{\ast} \left( {E} \right) = \mu ^{\ast} \left( {A
\cap E} \right) + \mu ^{\ast} \left( {\complement_{E} A \cap E}
\right).$

The Jordan and Lebesgue measures involve respective mappings (I)
and (m*) on spaces, which must be provided with $ \cap ,{\kern
1pt}$  $\cup{\kern 1pt}$ and ${\kern 1pt}\complement.{\kern 1pt}$
In spaces of the $\mathbb{R}^{\rm n}$ type, tessellation by balls
are involved (Bounias and Bonaly, 1996), which again demands a
distance to be available for the measure of diameters of
intervals.

A set of measure naught has been defined by Borel (1912) first as a linear
set (E) such that, given a number (e) as small as needed, all points of E
can be contained in intervals whose sum is lower than (e). Applying Borel
intervals imposes that appropriate embedding spaces are available for
allowing these intervals to exist

\bigskip

\noindent {2.1.2. Distances}

\bigskip

Following Borel, the length of an interval F = [a,b] is :
\begin{equation}
\label{eq1}  \rm  L(F) {\kern 1pt} = ( {b{\kern 1pt} {\kern 1pt} -
{\kern 1pt} {\kern 1pt} a} )\,\, - \,\sum\nolimits_{n} {L{\kern
1pt} {\kern 1pt} ( {C_{n}} )}
\end{equation}

\noindent where $\rm C_{n} $ are the adjoined, i.e. the open
intervals inserted in the fundamental segment. Since the Hausdorff
distances, as well as most of classical ones, is not necessarily
compatible with topological properties of the concerned spaces,
Borel provided an alternative definition of a set with measure
naught: the set (E) should be Vitali-covered by a sequence of
intervals ($\rm U_{n} $) such that: (i) each point of E belongs to
a infinite number of these intervals; (ii) the sum of the
diameters of these intervals is finite.

The intervals can be replaced by topological balls and therefore the
evaluation of their diameter still needs an appropriate general definition
of a distance.

A more general approach (Weisstein, 1999) involves a path $\rm
\varphi {\kern 1pt} {\kern 1pt} \left( {x,{\kern 1pt} y} \right)$
such that $\rm \varphi \left( {0} \right) = x$ and $\rm \varphi
{\kern 1pt} {\kern 1pt} \left( {1} \right) = y.$ For the case of
sets A and B in a partly ordered space, the symmetric difference
$\rm \Delta \left( {A,{\kern 1pt} B} \right) = C_{A \cup B} \left(
{A \cap B} \right)$ has been proved to be a true distance also
holding for more than two sets (Bounias and Bonaly, 1996; Bounias
1999). However, if $\rm A \cap B{\kern 1pt} {\kern 1pt} = {\kern
1pt} \O,$ this distance remains $\rm \Delta = A \cup B,$
regardless of the situation of A and B within an embedding space E
such that $\rm \left( {A,B} \right) \subset E.$ A solution to this
problem has been derived in (Bounias and Krasnoholovets, 2003a).

\bigskip

\noindent{2.1.3. Space Dimensions}

\bigskip

One important point is the following: in a given set of which
members structure is not previously known, a major problem is the
distinction between unordered N-uples and ordered N-uples. This is
essential for the assessment of the actual dimension of a space.

\textit{Fractal to Topological Dimension.} Given a fundamental
segment (AB) and intervals $\rm Li{\kern 1pt} {\kern 1pt} = {\kern
1pt} {\kern 1pt} \left[ {Ai,{\kern 1pt} {\kern 1pt} A\left( {i +
1} \right)} \right]$, a generator is composed of the union of
several such intervals: $\rm G{\kern 1pt} {\kern 1pt} = \cup
{\kern 1pt} _{\left( {{\kern 1pt} i{\kern 1pt} {\kern 1pt} \in
{\kern 1pt} {\kern 1pt} \left[ {1,n} \right]{\kern 1pt}} \right)}
{\kern 1pt} \left( {L{\kern 1pt} i} \right).$ Let the similarity
coefficients be defined for each interval by $\rm \rho {\kern 1pt}
i {\kern 1pt} {\kern 1pt} = {\kern 1pt} dist ( {Ai,A({i + 1})}) /
dist\left( {AB} \right).$

The similarity exponent of Bouligand is (e) such that for a generator with n
parts

\begin{equation}
\label{eq2} \rm  \sum\nolimits_{{\kern 1pt} _{\left( {i{\kern 1pt}
{\kern 1pt} \in {\kern 1pt} {\kern 1pt} \left[ {1,{\kern 1pt} n}
\right]} \right)}}  {\left( {\rho {\kern 1pt} i} \right)} ^{e} =
1.
\end{equation}

When all intervals have (at least nearly) the same size, then the various
dimension approaches according to Bouligand, Minkowsky, Hausdorff and
Besicovitch are reflected in the resulting relation :
  $$
 \rm   n \cdot \left( {\rho}  \right)^{e} = 1,
  \eqno(3\rm a)
  $$
that is,
  $$
 \rm e = Log {\kern 1pt} n{\kern 1pt}  / {\kern 1pt} Log {\kern 1pt} \rho.
 \eqno(3\rm b)
 $$
When e is an integer, it reflects a topological dimension, since
this means that a fundamental space E can be tessellated with an
entire number of identical balls B exhibiting a similarity with E,
upon coefficient $\rho .$

{\textit{Parts, Ordered N-uples and Simplexes}.} The following
structure is called a "descent"
  $$
 \rm E^{\left( {n} \right)} = \Big\{ e^{\left( {n}
\right)},{\kern 1pt} {\kern 1pt} {\kern 1pt} ( {E^{\left( {n + 1}
\right)}} ) \Big\}
   \eqno(4)
  $$

\noindent where (E) denotes a set or part of set, and (e) is a
kernel. A descent is finite if none of its parts is infinitely
iterated. A second kind set, F = \{a, (b, (c, (Z)))\}, is ordinary
if its descent is finite and extraordinary if its descents include
some infinite part. A fractal feature in an extraordinary second
kind set is recognizable.

Let the members of the above set E be ordered into a structure of
the type of F, e.g. $\rm {F}' = \{ a,\left( {b,\left( {c,\left(
{d,e,f} \right)} \right)} \right)\} $. Usually, a part (a,b) or
\{a,b\} is not ordered until it can be written in the form (ab) =
\{\{a\}, \{a,b\}\}. Stepping from a part (a,b,c,…) to a N-uple
(abc…) needs that singletons are available in replicates.

A simplex is the smaller collection of points that allows the set
to reach a maximum dimension. In a general acceptation, it should
be noted that the singletons of the set are called vertices and
ordered N-uples are (N-1)-faces $\rm A^{N - 1}$.

One question emerging now with respect to the purpose of this study is the
following: given a set of N points, how to evaluate the dimension of the
space embedding these points?

\subsection*{3. Distances and Dimensions Revisited}

\medskip

\subsubsection*{3.1. The Relativity of a General Form
                of Measure and Distance}

\medskip
\hspace*{\parindent} Our approach aims at searching for distances
that would be compatible with both the involved topologies and the
scanning of objects not yet known in the studied spaces. No such
configuration is believed to be an exception in a general case.

\bigskip

\noindent {3.1.1. General Case of a not Necessarily Ordered
Topological Distance}

\bigskip

\textbf{Proposition}. A generalized distance between spaces A,B
within their common embedding space E is provided by the
intersection of a path-set $\rm \varphi {\kern 1pt} ( {A,B} )$
joining each member of A to each member of B with the
complementary of A and B in E, such that $\rm \varphi {\kern 1pt}
({A,B})$ is a continued sequence of a function $f$ of a gauge (J)
belonging to the ultrafilter of topologies on \{E, A, B,...\}.

The path $\varphi {\kern 1pt} \left( {A,B} \right)$ is a set
composed as follows:

(i) $\rm \varphi {\kern 1pt} ( {A,B} ) = \bigcup\limits_{a \in
{\kern 1pt} {\kern 1pt} A,\,\,b{\kern 1pt} {\kern 1pt} \in B}
{\varphi {\kern 1pt} {\kern 1pt} \left( {a,b} \right)},$ {\kern
1pt} all defined on a sequence interval $\rm \left[ {0,{\it
f}^{n}\left( {x} \right)} \right],{\kern 1pt} {\kern 1pt} {\kern
1pt} {\kern 1pt} x \in E.$ The relative distance of A and B in E,
noted $\rm \Lambda _{E} {\kern 1pt} {\kern 1pt} \left( {A,B}
\right)$ is contained in $\rm \varphi {\kern 1pt} ( {A,B} )$:
  $$
  \rm \Lambda _{E} {\kern 1pt} {\kern 1pt} \left( {A,B} \right)
\subseteq \varphi {\kern 1pt} {\kern 1pt} \left( {A,B} \right).
    \eqno(5)
  $$

Denote by $\rm E^{ \circ} $ the interior of E, then
 $$\rm min\{ \varphi {\kern 1pt} {\kern 1pt} \left( {A,B} \right){\kern 1pt} \,
\cap E^{ \circ} \} {\kern 3pt} is {\kern3pt} a {\kern 3pt}
geodesic {\kern 3pt} of {\kern 3pt} space {\kern 3pt} E {\kern
3pt} connecting {\kern 3pt} A {\kern 3pt} to {\kern 3pt} B,
\eqno(6)
 $$
 $$\rm  max\{ \varphi {\kern 1pt} {\kern 1pt} \left( {A,B} \right)\,
 {\kern 1pt} \cap C_{{\kern 1pt} E^{ \circ} } \left( {A \cup B} \right)\}
  {\kern 3pt} is {\kern 3pt} a {\kern 3pt} tessellation {\kern 3pt} of
  {\kern 3pt} E {\kern 3pt} out {\kern 3pt} of {\kern 3pt} A {\kern 3pt}
  and {\kern 3pt} B. \eqno(7)
 $$
It is noteworthy that relation (6) refers to $\rm dim{\kern 1pt}
{\kern 1pt} \Lambda_{\rm E} = dim{\kern 1pt} \varphi,$ while in
relation (7) the dimension of the probe is that of the scanned
sets;

(ii) $\rm \varphi {\kern 1pt} {\kern 1pt} \left( {A,B}
\right)\,{\kern 1pt} \cap E^{ \circ} $ is a growing function
defined for any Jordan point, which is a characteristic of a
Gauge. In addition, the operator $\rm \Lambda _{{\kern 1pt}E}
{\kern 1pt} \left( {A,B} \right)$ meets the characteristics of a
Frechet metrics, since the proximity of two points $\rm
\underline{a}$ and $\rm \underline{b}$ can be mapped into the set
of natural integers and even to the set of rational numbers: for
that, it suffices that two members $\rm \varphi {\kern 1pt} \left(
{{\it f}^{n}\left( {x} \right),{\it f}^{n}\left( {y} \right)}
\right)$ are identified with a ordered pair $\rm \{ \varphi {\kern
1pt} \left( {{\it f}^{n}\left( {x} \right)} \right),\{ \varphi
{\kern 1pt} {\kern 1pt} \left( {{\it f}^{n}\left( {x} \right),{\it
f}^{n}\left( {y} \right)} \right)\} \} ;$

(iii) suppose that one path $\rm \varphi {\kern 1pt} ( {A,B})$
meets an empty space $\{ \O \}.$ Then a discontinuity occurs and
there exists some i, such that $\phi\left(f^{\rm i} ({\rm
\underline{b}})\right) = \O$. If all $\rm \varphi {\kern 1pt} (
{A,B} )$ meets $\{ \O \} ,$ then no distance is measurable. As a
corollary, for any singleton \{x\}, one has $\rm \varphi {\kern
1pt} {\kern 1pt} \left( {f\left( {\{ x\}} \right)} \right) = \O .$
The above properties meet other characteristics of a gauge: first,
given closed sets \{A,{\kern 1pt}B,{\kern 1pt}C, ...\}= E, then a
path set $\rm \varphi {\kern 1pt} \left( {E, {\kern 1pt}E \OE}
{\kern 1pt} \right)$ exploring the distance of E to the closure
$\left( {\OE} \right)$ of E meets only open subsets, so that $\rm
\varphi {\kern 1pt} \left( {E, {\kern 1pt}E \OE} {\kern 1pt}
\right)\,{\kern 1pt} = \O ;$ second, this is consistent with a
property of the Hausdorff distance:
  $$
  \rm max\{ \left( {A,B} \right) \subset E|\Lambda
_{E} \left( {A,B} \right)|\} \mapsto \quad diam_{{\kern 1pt}
Hausdorff} \left( {E} \right) \eqno(8)
  $$

\noindent
in all cases;

(iv) $\rm \Lambda _{E} \left( {A,B} \right)\, = \Lambda _{E}
\left( {B,A} \right)\,$ and $\rm \Lambda _{E} \left( {\{ x,y\}}
\right){\kern 1pt}  =  {\kern 1pt} \O \, \Leftrightarrow \,
x{\kern 1pt}  = {\kern 1pt} y.$ If the triangular inequality
condition is fulfilled, then $\rm \Lambda _{E} \left( {A,B}
\right)$ will meet all of the properties of a mathematical
distance.

\bigskip

\noindent{3.1.2. The Case of Topological Spaces}

\bigskip

 \textbf{Proposition.} A space can be subdivided in two
main classes: objects and distances.

The set-distance is the symmetric difference between sets: it has
been proved that it owns all the properties of a true distance
(Bounias and Bonaly, 1996) and that it can be extended to
manifolds of sets (Bounias, 1997). In a topologically closed
space, these distances are the open complementary of closed
intersections called the "instant". Since the intersection of
closed sets is closed and the intersection of sets with nonequal
dimensions is always closed, as was shown previously, the instants
rather stand for closed structures. Since the latter have been
shown to reflect physical-like properties, they denote objects.
Then, the distances as being their complementaries will constitute
the alternative class: thus, a physical-like space may be globally
subdivided into objects and distances as full components. The
properties of the set-distance allow an important theorem to be
now stated.

\textbf{Theorem.} Any topological space is metrizable as provided
with the set-distance $\left( {\Delta}  \right)$ as a natural
metrics. All topological spaces are kinds of metric spaces called
"delta-metric spaces".

The symmetric distance fulfills the triangular inequality,
including in its generalized form, it is empty if A = B = ..., and
it is always positive otherwise. It is symmetric for two sets and
commutative for more than two sets. Its norm is provided by the
following relation $\rm ||\Delta \left( {A} \right)|| = \Delta
\left( {A,\O}  \right).$

Therefore, any topology provides the set-distance which can be
called a topological distance and a topological space is always
provided with a self mapping of any of its parts into any one
metrics: thus any topological space is metrizable. Reciprocally,
given the set-distance, since it is constructed on the
complementary of the intersection of sets in their union, it is
compatible with existence of a topology. Thus a topological space
is always a "delta-metric" space.

Distance $\rm \Delta \left( {A,B} \right)$ is a kind of an
intrinsic case $\rm \left[ {\Lambda _{\left( {A,{\kern 1.5pt} B}
\right)} \left( {A,B} \right)} \right]$ of $\rm \Lambda _{E}
\left( {A,B} \right),$ while $\rm \Lambda _{E} \left( {A,B}
\right)$ is called a "separating distance". The separating
distance also stands for a topological metrics. Hence, if a
physical space is a topological space, it will always be
measurable. In other words, only in this case the physical space
can be ensured with a metrics.

\bigskip

\noindent{3.1.3. Dimensionality}

\bigskip

 A collection of scientific observation through experimental
devices produce images of some reality, and these images are
further mapped into mental images into the experimentalist's brain
(Bounias, 2000a). The information from the explored space thus
stands like parts of an apparatus being spread on the worker's
table, in view of a further reconstitution of the original object.
We propose to call this situation an informational display, likely
composed of elements with dimension lower than or equal to the
dimension of the real object.

In the construction of the set $\mathbb{N}{\kern 1pt} $ of natural
integers, von Neumann provided an equipotent from using replicates
of $\O :$
\[
0 = \O ,\,\,1 = \{ \O \} ,\,\,2 = \{ \O ,\{ \O \} \} ,\,\,3 = \{
\O ,\{ \O \} ,\{ \O ,\{ \O \} \} \} ,...
\]

A von Neumann set is Mirimanoff-first kind since it is isomorphic
with none of its parts. However this construction, associated with
the application of Morgan's laws to $\left( {\O}  \right),$
allowed the empty set to be attributed an infinite descent of
infinite descents and thus to be classified as a member of the
hypersets family (Bounias and Bonaly, 1997).

The following propositions can be proved:

(i) in an ordered pair $\rm \{ \{ x\} ,\{ x,y\} \} ,$ the paired
part $\rm \{ x,y\} $ is unordered;

(ii) an abstract set can be provided with at least two kinds of
orders: one with respect to identification of a max or a min, and
one with respect to ordered N-uples. These two order relations
become equivalent upon additional conditions on the nature of
involved singletons. Thus a fully information measure should
provide a picture of a space allowing its set component to be
depicted in terms of ordered N-uples $\rm \left( {N = 0 \to n, \ n
\subset {\kern 1pt} \mathbb{N}} \right).$

Regarding the definition of diameter (8), the following relation
can be proposed:
  $$
  \rm diam_{{\it f}} {\kern 1pt} {\kern 1pt} \left( {E}
\right) = \{ \left( {x,y} \right) \in E \ : \ max  {\it f}{\kern
1pt} ^{i}{\kern 1pt} \left( {Id\left( {x} \right)} \right) \cap
{\kern 1pt}  max {\it f}^{{\kern 1pt} i}{\kern 1pt} {\kern 1pt}
\left( {Id\left( {y} \right)} \right)\} \eqno(9)
  $$

\noindent
that gives a (N-2) members parameter (here E is a non-ordered set, Id the
identity self map of E, and $f$ the difference self map of E).

A set distance $\Delta$ is provided by the symmetric difference or
by a n-D Borel measure. A diameter is evaluated on E as the
following limit:
\begin{eqnarray} && \rm  diam\left( {E} \right)   \nonumber  \\
  && \rm = \{ inf{\kern 1pt} A \to
inf{\kern 1pt} E, {\kern 1pt} sup A \to min E, {\kern 1pt}
inf{\kern 1pt} B \to max {\kern 1pt} E, {\kern 1pt} sup {\kern
1pt} B \to sup {\kern 1pt} E {\kern 1pt} \Big|_{lim {\kern 2pt}
dist_{E} {\kern 1pt} ({A,{\kern 0.5pt}B})} \} . \nonumber (10)
 \end{eqnarray}

These approaches allow the measure of the size of tessellating balls as well
as that of tessellated spaces. A diagonal-like part of an abstract space can
be identified with and logically derived as, a diameter.

Each time a measure if obtained from a system, this means that no absolutely
empty part is present as an adjoined segment on the trajectory of the
exploring path. Thus, no space accessible to some sort of measure is
strictly empty in a both mathematical and physical sense, which supports the
validity of the quest of quantum mechanics for a structure of the void.

\bigskip

\noindent{3.1.4. The Dimension of an Abstract Space: Tessellating
With Simplex k-faces}

\bigskip

A major goal in physical exploration will be to discern among
detected objects which ones are equivalent with abstract ordered
N-uples within their embedding space. A first of further coming
problems is that in a space composed of members identified with
such abstract components, it may not be found tessellating balls
all having identical diameter. Thus a measure should be used as a
probe for the evaluation of the coefficient of size ratio $\left(
{\rho}  \right)$ needed for the calculation of a dimension.

The following principles should be hold:

(i) a 3-object has dimension 2 iff given $\rm A_{max}^{1} {\kern
1pt} $ its longer side fulfills the condition, for the triangular
strict inequality, where M denotes an appropriate measure
  $$
\rm  M\left( {A_{max}^{1}}  \right)\,{\kern 1pt} < {\kern 1pt}
M\left( {A_{2}^{1}}  \right) +  M\left( {A_{3}^{1}} \right);
  \eqno(11)
  $$

(ii) let space X be decomposed in into the union of balls
represented by D-faces $\rm A^{D}$ proved to have dimension $\rm
dim {\kern 1pt}\left( {A^{D}} \right)\, = \,D{\kern 1pt} $ by
relation (11) and size $\rm M\left( {A^{1}} \right){\kern 1pt} $
for a 1-face. Such a D-face is a D-simplex $\rm Sj$ whose size, as
a ball, is evaluated by $\rm M\left( {A_{max}^{1}}  \right)^{D} =
Sj^{D}.$ Let $\mathcal{N}$ the number of such balls that can be
filled in a space H, so that
  $$
 \rm \bigcup\limits_{i\, = 1}^{\mathcal{N}} {\{ S_j^{D}\}}
 \subseteq \left( {H \approx L_{max}^{d}}  \right)
 \eqno(12)
  $$
\noindent with H being identified with a ball whose size would be
evaluated by $\rm L^{d},$ L the size of a 1-face of H, and d the
dimension of H. Then, if $\rm \forall \,S_j, {\kern 2pt} S_j
\approx S_o,$ the dimension of H is:
  $$
\rm d\left( {H} \right) \approx \left( {D \cdot Log {\kern 1pt}
S_o +  Log {\kern 1pt} \mathcal{N}} \right){\kern 1pt} /{\kern
1pt} Log\,L_{max}^{1} .
 \eqno(13)
  $$

Relation (13) stands for a kind of interior measure in the
Jordan's sense. In contrast, if one poses that the reunion of
balls covers the space H, then, d(H) rather represents the
capacity dimension, which remains an evaluation of a fractal
property.

In the analysis of a abstract space $\rm H = L^{x},$ of which
dimension x is unknown, the identification of which members can be
identified with N-uples supposedly coming from a putative
Cartesian product of members of H, e.g. $\rm G \subset
H\,:\,\angle \,a \in G, \quad \left( {aaa...a} \right)_{h} \in
G^{h},$ is allowed by an anticipatory process (because by
definition the later enables the construction of the power set of
parts (${\cal P}^{\rm h}$) of G).

\subsection*{4. A Universe as a Mathematical Lattice}

\subsubsection*{4.1. Defining a Probationary Space}

\hspace*{\parindent} The universe from nothing. How is it
possible? The question is in the center of attention of Marcer at
al. (2003). In our works, we have looked at the problem starting
from an idea of a probationary space. The probationary space
(Bounias, 2001) is defined as a space fulfilling exactly the
conditions required for a property to hold, in terms of: (i)
identification of set components (ii) identification of
combinations of rules (iii) identification of the reasoning
system. All these components are necessary to provide the whole
system with decidability. Since lack of mathematical decidability
inevitably flaws also any physical model derived from a
mathematical background, our aim is to access as close as possible
to these imposed conditions in a description of both a possible
and a knowable universe, in order to refine in consequence some
current physical postulates.

The above considerations have raised a set of conditions needed
for some knowledge to be gettable about a previously unknown
space, upon pathwise exploration from a perceiving system A to a
target space B within an embedding space B. In the absence of
preliminary postulate about existence of so-called "matter" and
related concepts, it has been demonstrated that the existence of
the empty set is a necessary and sufficient condition for the
existence of abstract mathematical spaces $\rm \left( {W^{n}}
\right)$ endowed with topological dimensions (n) as great as
needed (Bounias and Bonaly, 1997). Hence, the empty set appears as
a set without members though containing empty parts. Indeed,
members do not appear, because they are nothing ($\O$) and at the
same time they are empty elements $\O$.

\subsubsection*{4.2. The Founding Element}

\hspace*{\parindent} It is generally assumed that some set does
exist. This strong postulate has been reduced to a weaker form
reduced to the axiom of the existence of the empty set (Bounias
and Bonaly, 1997). It has been shown that providing the empty set
$\rm \left( {\O}  \right)$ with $\left( { \in ,\,\, \subset}
\right)$ as the combination rules, that is also with the property
of complementarity ($\complement$), results in the definition of a
magma allowing a consistent application of the first Morgan's law
without violating the axiom of foundation iff the empty set is
seen as a hyperset, that is a nonwellfounded set. A further
support of this conclusion emerges from the fact that several
paradoxes or inconsistencies about the empty set properties are
solved (Bounias and Bonaly, 1997).

\subsubsection*{4.3. The Founding Lattice}

\hspace*{\parindent} \textbf{Theorem and Lemmas.} The magma $\rm
{\O} ^{\,{\O}}  = \{ {\O}, \complement \} $ constructed with the
empty hyperset and the axiom of availability is a fractal lattice
(writing ${\O} ^{\,{\O}} $ denotes that the magma reflects the set
of all self-mappings of $\rm \left( {\O} \right),$ which
emphasizes the forthcoming results).

The space constructed with the empty set cells of $E_{\O}  $ is a
Boolean lattice. Indeed, let $\rm  \cup \left( {\O}  \right) = S$
denote a simple partition of $\left( {\O}  \right).$ The
combination rules $ \cup $ and $ \cap $ provided with
commutativity, associativity and absorption are holding. In
effect: ${\O} \cup {\O} = {\O} ,$ ${\O} \cap {\O} = {\O} $ and
thus necessarily ${\O} \cup \left( {{\O} \cap {\O}} \right) = {\O}
, \quad {\O} \cap \left( {{\O} \cup {\O}} \right) = {\O}.$ Thus
space $\rm P\{ \left( {\O} \right),\,\left( { \cup ,{\kern 1pt}
{\kern 1pt} \cap} \right)\} $ is a lattice. The null member is
${\O} $ and the universal member is $2^{\O} $ that should be
denoted by $\aleph _{{\kern 1pt} {\O}}.$ Since in addition, by
founding property $\complement_{{\kern 1pt} {\O}} \left( {\O}
\right) = {\O} ,$ and the space of $\left( {\O} \right)$ is
distributive, then $\rm S\left( {\O} \right)$ is a Boolean
lattice.

 $\rm S\left( {\O}  \right)$ is provided with a topology of discrete space.
The lattice $\rm S\left( {\O}  \right)$ owns a topology. The space
$\rm S\left( {\O}  \right)$ is Hausdorff separated. Units $\left(
{\O} \right)$ formed with parts thus constitute a topology $\left(
{\mathcal{T}_{\O} }  \right)$ of discrete space.

The magma of empty hyperset is endowed with self-similar ratios.
The von Neumann notation associated with the axiom of
availability, applying on $\left( {\O}  \right),$ provides
existence of sets $\rm \left( {N^{\O} } \right)$ and $\rm \left(
{Q^{\O} } \right)$ equipotent to the natural and the rational
numbers (Bounias and Bonaly, 1997). Let us consider a Cartesian
product $\rm E_n \times E_n$ of a section of $\rm \left( {Q^{\O} }
\right)$ of n integers. The amplitude of the available intervals
range from 0 to n, with two particular cases: interval [0,{\kern
1pt} 1] and any of the minimal intervals [1/n-1, 1/n]. Interval
[0, 1/n{\kern 1pt}(n - 1)] is contained in [0,{\kern 1pt} 1].
Hence, since empty sets constitute the founding cells of the
lattice $S\left( {\O} \right),$ the lattice is tessellated with
cells (or balls) with homothetic-like ratios of at least r=
n{\kern 1pt}(n - 1). The absence of unfilled areas is supported by
introduction of the "set with no parts" (Bounias and
Krasnoholovets, 2003b).

Such a lattice of tessellation balls has been called the "tessellattice"
(Bounias and Krasnoholovets, 2003a).

The magma of empty hyperset is a fractal tessellattice. Indeed,
$\rm \left( {\O}  \right) \cup \left( {\O}  \right){\kern 1pt}
{\kern 1pt} = {\kern 1pt} {\kern 1pt} \left( {{\O} ,{\O}} \right)$
and $\left( {\O}  \right) \cap \left( {\O} \right){\kern 1pt}
{\kern 1pt} = {\kern 1pt} {\O} ,$ moreover, the magma $\left(
{{\O} ^{\O} } \right) = \{ {\O} , \complement \} $ represents the
generator of the final structure, since $\left( {\O} \right)$ acts
as the "initiator polygon", and complementarity as the rule of
construction. These three properties stand for the major features
that characterize a fractal object. Finally, the axiom of the
existence of the empty set, added with the axiom of availability
in turn provide existence to a lattice $\rm S\left( {\O} \right),$
which constitutes a discrete fractal Hausdorff space.

\subsubsection*{4.4. Existence and Nature of Space-time}

\hspace*{\parindent} A lattice of empty sets can provide existence
to at least a physical-like space.

Let ${\O} $ denote the empty set as a case of the whole structure
and $\{ {\O} \} $ denotes some of its parts. The set of parts of
$\O $ contains parts equipotent to sets of integers, of rational
and of real numbers, and owns the power of continuum. Then,
looking at the infering spaces $\rm \left( {W^{n}}
\right),\;\left( {W^{m}} \right), \ ...$, we obtain
  $$
 \rm \{ \left( {W^{n}} \right) \cap \left( {W^{m}} \right)\}
 _{m{\kern 1pt}{\kern 1pt} > {\kern 1pt} {\kern 1pt} n} =
 \left( {\Theta ^{n}} \right), {\kern 3pt} which {\kern 3pt}
 is {\kern 3pt} a {\kern 3pt} closed {\kern 3pt} space.
  \eqno(14)
  $$

These spaces provide collections of discrete manifolds whose
interior is endowed with the power of continuum. Consider a
particular case $\left( {\Theta ^{4}} \right)$ and the set of its
parts $\rm {\cal{P}} \left( {\Theta ^{n}} \right);$ then any of
intersections of subspaces $\rm \left( {E^{d}} \right)_{{\kern
1pt} d{\kern 1pt} <  {\kern 1pt} 4} $ provides a d-space in which
the Jordan-Veblen theorem allows closed members to get the status
of both invesigated objects and perceiving objects. This stands
for observability, which is a condition for a space to be in some
sort observable, i.e., physical-like.

In any $\left( {\Theta ^{4}} \right)$ space, the ordered sequences
of closed intersections $\rm \{ \left( {E^{d}} \right)_{{\kern
1pt} d{\kern 1pt} {\kern 1pt} < {\kern 1pt} {\kern 1pt} 4} \},$
with respect to mappings of members of $\rm \{ \left( {E^{d}}
\right){\kern 1pt} _{d{\kern 1pt} {\kern 1pt} {\kern 1pt} < {\kern
1pt} {\kern 1pt} {\kern 1pt} 4} \} _{i} $ into $\rm \{ \left(
{E^{d}} \right)_{{\kern 1pt} d{\kern 1pt} {\kern 1pt} < {\kern
1pt} {\kern 1pt} 4} \} _{j},$ provides an orientation accounting
for the physical arrow of time (Bounias, 2000). Thus the following
proposition: A manifold of potential physical universes is
provided by the $\rm \left( {\Theta ^{n}} \right)$ category of
closed spaces.

Our space-time is one of the mathematically optimum ones, together
with the alternative series of $\rm \{ \left( {W^{3}} \right) \cap
\left( {W^{m}} \right)\} _{m{\kern 1pt} {\kern 1pt} > {\kern 1pt}
{\kern 1pt} 3} $.

\subsection*{5. Principles of Construction of Physical Space}

\hspace*{\parindent} Is space independent from matter or matter is
deformation of space? These discrepancies essentially come from
the fact that probationary spaces supporting a number of explicit
or implicit assessments have not been clearly identified.

At cosmological scales, the relativity theory places referential
in an undefined space, with undefined gauges for substrate for the
transfer of information and the support of interactions. Here,
distances are postulated without reference to objects: the
geometry of space is separated from matter (though the matter is
able to influence the space).

At quantum scales, a probability that objects are present in a certain
volume is calculated. But again, nothing is assessed about what are these
objects, and what is their embedding medium in which such ``volumes'' can be
found. Furthermore, whether these objects are of a nature similar or
different to the nature of their embedding medium has not been addressed. In
this case, objects are postulated without reference to distances.

We have treated some founding principles about the definition of the space
of magmas (i.e., the sets, combination rules and structures) in which a
given proposition can be valid. Such a space, when identified, is called a
probationary space (Bounias, 2001). Here, it will be presented the formalism
which leads from existence of abstract (e.g. purely mathematical) spaces to
the justification of a distinction between parts of a physical space, which
can be said empty, and parts that can be considered as filled with
particles. This question is thus dealing with a possible origin of matter
and its distribution, and changes in this distribution gives raise with
motion, i.e., with physics.

The values of the constants of electromagnetic, weak and strong
interactions as functions of distances between interacting
particles converge to the same at a scale about 10$^{-30}$ m. We
may assume that it is this point at which the cleavage of space
takes place, or in other words, the given scale corresponds to a
violation of space homogeneity. The model proceeds from the
assumption that all quantum theories (quantum mechanics,
electrodynamics, chromodynamics, etc.) are in fact only
phenomenological. Accordingly, for the understanding processes
occurring in the real microworld, one needs a submicroscopic
approach that in turn should be available for all peculiarities of
the microstructures of the real space. In other terms, gauges for
the analysis of all components of the observable universe should
belong to an ultrafilter (Bounias and Krasnoholovets, 2003a). The
study about the model of inertons (Krasnoholovets, 1997, 2002,
2004) has suggested that a founding cellular structure of space
shares discrete and continuous properties, which is also shown to
be consistent with the abstract theory of the foundations of
existence of a physical space (Bounias and Bonaly, 1997).

\subsubsection*{5.1. Foundations of Space-time}

\hspace*{\parindent} A possible application of fractal geometry to
the description of space-time has already been demonstrated by
Nottale (1997, 2003). He studied relativity in terms of fractality
basing his research on the Mandelbrot's concept of fractal
geometry. Nottale introduced a scale-relativity formalism, which
allowed him to propose a special quantization of the universe. In
his theory, scale-relativity is derived from applications of
fractals introduced as follows. The fractal dimension D is defined
from the variation with resolution of the main fractal variable,
i.e., the length $l$ of fractal curve plays a role of a fractal
curvilinear coordinate. He also introduced the topological
dimension D$_T$ determining it as ${\rm D}_T = 1$ for a curve, 2
for a surface, etc. The scale dimension then was determined as
$\delta = {\rm D} - {\rm D}_T$. If $\delta$ is constant, the above
relationship gives a power-law resolution dependence $l
=l_0(\ell/\epsilon)^\delta$. Such a simple scale-invariant law was
identified with a Galilean's kind of scale-relativistic law.

Generally speaking, Nottale's approach leads to the conclusion
that a trajectory of any physical system diverges due to the inner
stochastic nature that is caused by the fractal laws. In this
approach fractality is associated with the length of a curve as
such.

In the present work we show that the fractal geometry can be
derived from complete other mathematical principles, which becomes
possible on the basis of reconsidered fundamental notions of
space, measure, and length, which allow us to introduce deeper
first principles for the foundation of fractal geometry.

Our approach follows a hypothesis (Bonaly, 1992) that a
characteristic of a physical space is that it should be in some
way observable. This implies that an object called the "observer"
should be able to interact with other objects, "observed". We will
call them the "perceiver" and the "perceived" objects,
respectively. Bonaly's conjecture implied that perceived objects
should be topologically closed, otherwise they would offer no
frontier to allow a probe to reflect their shape. The existence of
closed topological structures and a proof was given that the
intersection of two spaces having nonequal dimensions owns its
accumulation points and is therefore closed. In other words, the
intersection of two connected spaces with nonequal dimensions is
topologically closed. This allowed the representation of the
fundamental metrics of space-time by a convolution product where
the embedding part
  $$
 {\rm U}4{\kern 1pt}  =  \int \Big( \int_{\rm dS} {\kern 1pt}
( {\rm d} {\vec {\rm x}}{\kern 2pt} {\rm d} {\vec {\rm y}} {\kern
2pt} {\rm d} {\vec {\rm z}} {\kern 2pt} ) {\kern 1pt} \Big) \ast
{\rm d} \psi ( {\rm w} )
  \eqno(15)
  $$

\noindent where dS is the element of space-time and $\rm d\psi
\left( {w} \right)$ the function accounting for the extension of
3-D coordinates to the 4$^{\rm th}$ dimension through convolution
$\left( { \ast}  \right)$ with the volume of space.

\bigskip

\noindent{5.1.1. Space-time as a Topologically Discrete Structure}

\bigskip

The mapping of two Poincar${\rm \acute{e}}$ sections is assessed
by using a natural metrics of topological spaces. Let $ \rm \Delta
\left( {A,B,C,...} \right)$ the generalized set distance as the
extended symmetric difference of a family of closed spaces,
   $$
\rm \Delta(A_{i})_{{\kern 1pt}i {\kern 1pt} \in {\kern 1pt}
N}={\mathop \complement_{\cup \{ Ai \}}}{\kern 1pt} {\mathop \cup
_{{\kern 2pt} i\neq j}}{\kern 1pt}(A_i \cap A_j).
   \eqno(16)
   $$

The complementary of $\Delta $ in a closed space is closed. It is
also closed even if it involves open components with nonequal
dimensions. In this system $\rm m {\kern 1pt} \langle \{ A_{i} \}
\rangle = \Delta $ has been associated with the instant, i.e., the
state of objects in a timeless  Poincar${\rm \acute{e}}$ section
(Bounias, 1997). Since distances $\Delta $ are the complementaries
of objects, the system stands as a manifold of open and closed
subparts.

The set-distance provides a set with the finer topology and the set-distance
of nonidentical parts provides a set with an ultrafilter. Regarding a
topology or a filter founded on any additional property $\left( { \bot}
\right),$ this property is not necessarily provided to a $\Delta $-filter.
The topology and filter induced by $\Delta $ are thus respectively the finer
topology and an ultrafilter.

The mappings of both distances and instants from one to another
section can be described by a function called the "moment of
junction", since it has the global structure of a momentum. Let an
indicatrix function 1(x) be defined by the correspondence of x
with some $\rm c\left( {x} \right)$ in $\rm S\left( {i{\kern 1pt}
+ 1} \right),$ i.e. 1(x) = 0 or 1. Then a function $\rm {\it
f}_{{\kern 1pt} \left( {E_{i},  {\kern 1pt} E_{i + 1}} \right)},$
or shortly $f_{\rm E}$ accounts for a distribution of the
indicatrix functions of all points out of the maximum number of
possibilities, which would be $\rm 2^{E}$ for the set of pairs of
set E (see Bounias and Krasnoholovets, 2003b). The proportion of
points involved in the mappings of parts of E{\kern
0.4pt}{\footnotesize{i}} into $\rm E\left( {i + 1} \right)$ equals
$\rm {\it f}_{{\kern 1pt}E}^{{\kern 1pt} E} \left( {E} \right) =
{\it f}_{{\kern 1pt} E} \left( {E} \right)/2^{E}, \quad 0 < {\it
f}_{{\kern 1pt}E}^{{\kern 1pt} E} \left( {E} \right) < 1.$

Two species of the moment of junction are represented by the
composition $\left( { \bot}  \right)$ of $\rm {\it f}_{E}^{E}
\left( {E} \right)$ with either the set-distance of the instant.
Hence
   $$
  \rm  MJ_{\Delta}  = \Delta \left( {E} \right){\kern 2pt} \bot
  {\kern 2pt}{\it f}_
  {{\kern 1pt}E}^{{\kern 1pt}E}
\left( {E} \right),
  \eqno(17{\rm a})
   $$
   $$
\rm  MJ_{m} = m{\kern 1pt} {\kern 1pt} \langle E\rangle {\kern
2pt} \bot {\kern 2pt} {\it f}_{{\kern 1pt}E}^{{\kern 1pt}E} \left(
{E} \right).
  \eqno(17{\rm b})
   $$

Generally, $\rm MJ_{\Delta}  \ne MJ_{m}.$ As a composition of
variables with their distribution, (17a,b) actually represent a
form of momentum.

\bigskip

\noindent{5.1.2. Space-time as Fulfilling a Nonlinear Convolution
Relation}

\bigskip

The "moments of junctions" (MJ) mapping an instant (a 3-D section
of the embedding 4-space) to the next one apply to both the open
(the distances) and their complementaries the closed (the
reference objects) in the embedding spaces. But points standing
for physical objects able to move in a physical space may be
contained in both of there reference structures.

Then, it appears that two kinds of mappings are composed with one
another.

A 'space-time'-like sequence of Poincar${\rm \acute{e}}$ sections
is a nonlinear convolution of morphisms. The demonstration
involves two kinds of mappings:

(i) mapping $\left( {\mathcal{M}} \right)$ connects a frame of
reference to the next one: here, the same organization of the
reference frame-spaces must be found in two consecutive instants
of our spacetime, otherwise, no change in the position of the
contained objects could be correctly characterized. However, there
may be some deformations of the sequence of reference frames.
Mappings $\left( {\mathcal{M}} \right)$ denote the corresponding
category of morphisms;

(ii) mapping $\left( {\mathcal{J}} \right)$ connects the objects of one
reference cell to the corresponding next one. Mappings $\left( {\mathcal{J}}
\right)$ thus behave as indicatrix functions of the situation of objects
within the frames.

Each section, or timeless instant of our space-time, is described
by a composition $\left( { \circ}  \right)$ of these two kinds of
morphisms: "space-time instant" = $\left( {\mathcal{M} \circ
\mathcal{J}} \right).$ Besides, stepping from one to the next
instant is finally represented by a mapping $T$, such that the
composition $\left( {\mathcal{M} \circ \mathcal{J}} \right)$ at
iterate (k) is mapped into a composition $\left( {\mathcal{M} \bot
\mathcal{J}} \right)$ at iterate (k+i): $\rm \left( {\mathcal{M}
\bot \mathcal{J}} \right)_{k + i} = {\kern 1pt} \mathbf{T}^{ \bot
}\left( {\mathcal{M} \circ \mathcal{J}} \right)_{k} .$ Hence,
mapping $\left( {\mathbf{T}^{ \bot} } \right)$ appears like a
relation that maps a function $\rm F_{i + k} $ into $\rm {F}'_{j +
k}.$ Such a relation represent a case of the general convolution,
which is a nonlinear and multidimensional form of the convolution
product. Mapping for the case of a integrable space gives
  $$
\rm {F}'\left( {X} \right)\, = \,\int {\alpha} {\kern 1pt} {\kern
1pt} \left( {{X}' - X} \right)\,{\kern 1pt} F\left( {X}
\right)\,{\kern 1pt} d{\kern 1pt} {X}'.
  \eqno(18)
  $$

\noindent This relation exhibits a great similarity with a
distribution of functions in the Schwartz sense or a convolution
product
  $$
\rm \int_{E} {\it f} ( X{\kern 1pt} {\kern 1pt} - {\kern 1pt}
{\kern 1pt} u ) F ( {u} ) d{\kern 1pt} ( {u}) = ( {\it f} \ast
 F )( {X} ).
  \eqno(19)
  $$

Thus the connection from the abstract universe of mathematical
spaces and the physical universe of our observable space-time is
provided by a convolution of morphisms, which supports the
conjecture of relation (15).

\subsubsection*{5.2. Relative Scales in the Empty-set Lattice}

\noindent{5.2.1. Quantum Levels at Relative Scales}

\bigskip
Inside any of the above spaces, properties at microscale are
provided by properties of the spaces whose members are empty set
units.

Particular levels of a measure of these units can be discerned. It
has been demonstrated (Bounias and Krasnoholovets, 2003b) that the
Cartesian product of a finite beginning section of the integer
numbers provides a variety of nonequal empty intervals. This means
that a finite set of rational numbers inferring from a Cartesian
product of a finite beginning section of integer numbers
establishes a discrete scale of relative sizes. Indeed, intervals
are constructed from the corresponding mappings. For example, with
D=2, the smaller ratios available are 1/n and $\rm 1/(n - 1)$, so
that their distance is the smaller interval is $\rm 1/n(n-1)$.
Smaller intervals $\left( {\sigma}  \right)$ in $\rm \left(
{E_{n}} \right)^{3}$ obey the order of increasing sizes:
\begin{eqnarray}
&&    \rm   \forall {\kern 1pt}n > 1 :     \nonumber   \\ && \rm (
{\sigma _{( {i} )}} )\, = 1/n^{2}( {n - 1} ) < ( {\sigma _{( {ii}
)}} )\, = 1/n( {n - 1} )^{2} < ( {\sigma _{( {iii} )}} ) = {n - 1}
){\kern 1pt} /{\kern 1pt} n^{2} ( {n - 1} ).   \nonumber
  \quad (20)
  \end{eqnarray}

The maximal of the ratios of larger (n) to smaller $\rm \left(
{1/n^{2}\left( {n - 1} \right)} \right)$ segments is
  $$
\rm max{\kern 1pt} {\kern 1pt} \left( {\sigma}  \right){\kern 1pt}
{\kern 1pt} /{\kern 1pt} min{\kern 1pt} {\kern 1pt} \left(
{\sigma}  \right) = n^{3}\left( {n{\kern 1pt} {\kern 1pt} - 1}
\right). \eqno(21)
  $$

A scaling progression covering integer subdivisions (n) divides a
fundamental segment ($\rm n=1$) by 2, then each subsegment by 3,
etc. Therefore the size of structures is a function of iterations
(n). At each step $\rm \left( {\nu_j} \right)$ the ratio of size
in dimension D is $\rm \left( {\Pi \nu_j} \right)^{D},$ so that
the maximal, as follows from (21), is
  $$
\rm \rho \propto \{ ( {\Pi \nu_j} )^{D} ( ( {\Pi \nu_j})
 - 1 \} _{j{\kern 1pt} \linebreak = {\kern 1pt} 1  \to  {\kern 1pt} n}.
  \eqno(22)
 $$

Values (22) can be written as $\rm \rho_j = a_j.{\kern
1pt}10^{x_j}$ where in base 10 one takes $\rm a_j.$ belonging to
the neighborhood of unity, i.e., $\rm a_j. \in {\kern 1pt} ]1[$,
and look at the corresponding integer exponents $\rm x_j$ as the
order of sizes of structures constructed from the lattice $\rm
\ell = \left( {\Pi \nu_j} \right)^{D}.$ Regarding distances $\rm
\left( {D{\kern 1pt} {\kern 1pt} {\kern 1pt} = {\kern 1pt} {\kern
1pt} 1} \right)$ to areas and volumes ($\rm D{\kern 1pt} = {\kern
1pt} 2$ and 3), Eq. (22) consistently provides the orders of
physical scales, from microscopic to cosmic, as has been
demonstrated by Bounias and Krasnoholovets (2003b).

\subsection*{6. Matter Generated in a Lattice Universe}

\hspace*{\parindent} Space represented by the lattice $\rm F\left(
{U} \right){\kern 1pt} \cup {\kern 1pt} \left( {W} \right){\kern
1pt} \cup ({\not{\!}{c}\!\!}{\kern 3pt})$, where $\rm
({\not{\!}{c}\!\!} {\kern 3pt})$ is the set with neither members
nor parts, accounts for relativistic space and quantic void,
because (i) the concept of distance and the concept of time have
been defined on it and (ii) this space holds for a quantum void
since it provides a discrete topology, with quantum scales and it
contains no "solid" object that would stand for a given provision
of physical matter.

The sequence of mappings of one into another structure of reference (e.g.
elementary cells) represents an oscillation of any cell volume along the
arrow of physical time.

However, there is a transformation of a cell involving some iterated
internal similarity, which precludes the conservation of homeomorphisms. If
N similar figures with similarity ratios 1/r are obtained, the Bouligand
exponent (e) is given by
  $$
\rm  N \cdot \left( {1/r} \right){\kern 1pt} ^{e}{\kern 1pt}
{\kern 1pt} {\kern 1pt} = {\kern 1pt} {\kern 1pt} {\kern 1pt} 1
 \eqno(23)
  $$

\noindent and the image cell gets a dimensional change from d to
$\rm {d}'{\kern 1pt} {\kern 1pt} = {\kern 1pt} {\kern 1pt} {\kern
1pt} \ln ( {N} ){\kern 1pt} {\kern 1pt} /{\kern 1pt} {\kern 1pt}
\ln( {{\kern 1pt} r} ) = e > 1.$ In this case the putatively
homeomorphic part of the image cell is no longer a continued
figure and the transformed cell no longer owns the property of a
reference cell.

This transformation stands for the formation of a "particle" also
called "particled cell", or more appropriately "particled ball",
because it is a kind of topological ball ${\rm B}\left[ {\O
,{\kern 1pt} {\kern 1pt} r\left( {\O}  \right)} \right].$ Thus a
particled ball is represented by a nonhomeomorphic transformation
in a continuous deformation of space elementary cells.

\subsubsection*{6.1. Quanta of Fractality}

\hspace*{\parindent} A minimum fractal structure is provided by a
self-similar figure whose combination rule includes an initiator
and a generator, and for which the similarity dimension exponent
is higher than unity.

Let an initial figure (A) be subdivided into r subfigures at the
first iteration and let (r+a) be the number of subfigures
constructed on the original one. Since each i$^{\rm th}$ iteration
involves the subvolume $\rm v_i$, in the simplest case we
anticipate $\rm v_i= v_{\rm i-1}{\kern 1pt}\cdot(1/r)^3$. A
fractal decomposition consists in the distribution of the members
of the set of fractal subfigures $\rm \Gamma \supset \{
\sum\nolimits_{\left( {{\kern 1pt} i =  1 {\kern 1pt} \to {\kern
1pt} \infty}  \right)} {\{ \left( {r{\kern 1pt} {\kern 1pt} +
{\kern 1pt} a} \right)^{i} \cdot v_{i{\kern 1pt} - 1} \cdot \left(
{1/r} \right)^{3}\}}  \} $ constructed on one figure, among a
number of connected figures (C$_{1}$, C$_{2}$, ..., C$_{\rm k}$)
similar to the initial figure (A). If k reaches infinity, then all
subfigures of A are distributed and (A) is no longer a fractal.

\subsubsection*{6.2. Interactions Involving Exchanges of
Structures}

\hspace*{\parindent} The motion of any particle must be
accompanied by the corresponding cloud of space excitations, this
is the major thesis of submicroscopic mechanics (Krasnoholovets,
1997,{\kern 2pt}2002,{\kern 2pt}2003). These excitations called
"inertons" appear due to the friction of the moving particle,
which the latter experiences at its motion in the tessellattice
tightly packed with cells (or balls, or superparticles).

Let a ball (A) containing a fractal subpart on it. Deformations can be
transferred from one to another ball with conservation of the total volume
of the full lattice (which is constituted by a higher scale empty set). If a
fractal deformation is subjected to motion, it will collide with surrounding
degenerate balls. Such collisions will result in fractal decompositions at
the expense of (A) whose exponent (e$_{A}$) will decrease, and to the profit
of degenerate cells. This is a typical scattering from friction.

The remaining of fractality decreases from the kernel (i.e. the
area adjacent to the original particled deformation) to the edge
of the inerton cloud. At the edge, depending on the local
resistance of the lattice, the decomposition (denoted as the
n$^{\rm th}$ iteration) can result in $\rm \left( {e_{n}}  \right)
= 1.$ Therefore, while central inertons exhibit decreasing higher
boundaries, edge inertons are bounded by a rupture of the
remaining fractality.

\subsubsection*{6.3. Mass}

\hspace*{\parindent} A particled ball as described above provides
formalism describing the elementary particles proposed by
Krasnoholovets (1997; 2000). In this respect, mass is represented
by a fractal reduction of volume of a ball, while just a reduction
of volume as in degenerate cells is not sufficient to provide
mass. The mass $\rm m_{A} $ of a particled ball A is a function of
the fractal-related decrease of the volume of the ball
  $$
\rm  m_{A} \propto \left( {1/V^{part}} \right) \cdot \left(
{e_{\nu} - {\kern 1pt} {\kern 1pt} 1} \right){\kern 1pt} _{e_{\nu}
{\kern 1pt} \geqslant {\kern 1pt} 1}
  \eqno(24)
  $$

\noindent where (e) is the Bouligand exponent, and (e-1) the gain
in dimensionality given by the fractal iteration; the index $\nu$
denotes the possibly fractal concavities affecting the particled
ball. Just a volume decrease is not sufficient for providing a
ball with mass, since a dimensional increase is a necessary
condition.

\subsubsection*{6.4. Particles and Inertons}

\hspace*{\parindent} Two interaction phenomena have been
considered (Bounias and Krasnoholovets, 2003b): first, the
elasticity $\left( {\gamma}  \right)$ of the lattice favours an
exchange of fragments of the fractal structure between the
particled ball and the surrounding degenerate balls. In a first
approach, the resulting oscillation has been considered
homogeneous. Second, if the particled ball has been given a
velocity, its fractal deformations collide with neighbour
degenerate balls and exchanges of fractal fragments occur.

The velocity of the transfer of deformations is faster for
non-fractal deformations and slower for fractal ones, at slowering
rates varying as the residual fractal exponent $\rm \left( e_{i}
\right).$ The motion of the system constituted by a particled ball
and its inerton cloud provides the basis for the de Broglie and
Compton wavelength.

The system composed with the particle and its inertons cloud is not likely
to be of homogeneous shape. This property will be accounted for in
spin-related properties of observable matter (see also Krasnoholovets,
2000).

The fractality of particle-giving deformations gathers its space
parameters $\rm ( \varphi_ {\kern 1pt i} )_i$ and velocities $\rm
( v )$ into a self-similarity expression that provides a
space-to-time connection. Indeed, let $( \varphi _{ \circ})$ and
$\rm ( v_{ \circ} )$ be the reference values. Then the similarity
ratios are $\rm \rho ( \varphi ) = ( \varphi_ {\kern 1pt i})/(
\varphi _{ \circ} )$ and $\rm \rho ( {v} ) = v/v_{ \circ},$
therefore,
   $$
\rm \rho {\kern 1pt} {\kern 1pt} \left( {\varphi} \right)^{e} +
\rho {\kern 1pt} \left( {v} \right)^{e} = 1.
  \eqno(25)
   $$
Once again, the right hand side of eq. (25) includes only space
and space-time parameters.

Then for distances $\rm \left( {l} \right)$ and masses $\left( {m}
\right)$ we obtain using equation (25)
   $$
\rm  \left( {l/l_{{\kern 1pt} \circ} } \right)^{2} + \left(
{v{\kern 1pt} /{\kern 1pt} v_{ \circ} } \right)^{2} = 1\quad
\Leftrightarrow \quad l = l_{{\kern 1pt} \circ}  \cdot \sqrt{
1{\kern 1pt}  -  ( {v / v_{ \circ} })^{2} },
  \eqno(26)
  $$
  $$
\rm  \left( {{\it m}_{{\kern 1pt} \circ}  /{\it m}} \right)^{2} +
\left( {v{\kern 1pt} /{\kern 1pt} v_{ \circ} } \right)^{2} =
1\quad \Leftrightarrow \quad {\it m} = {\it m}_{{\kern 1pt} \circ}
/ \sqrt{ 1{\kern 1pt} - ( {v} /v_{ \circ } )^{2})},
  \eqno(27)
  $$

The Lagrangian $\left( {\mathcal{L}} \right)$ should obey a
similar law and $\left( {\mathcal{L}/\mathcal{L}_{\, \circ} }
\right)$ should fulfill relation (25) as a form of $\rm \rho
{\kern 1pt} \left( {\varphi} \right)^{e}$. Then, $\rm \left(
{\mathcal{L}/\mathcal{L}_{\circ} } \right)^{2} + \left( {v{\kern
1pt} /{\kern 1pt} v_{ \circ} }  \right)^{2} = 1$ and analogously
we can take $\rm \mathcal{L}_{\, \circ}  = - {\it m}{\kern 1pt}
v_{ \circ} ^{2} ,$ thus finally $\rm \mathcal{L} = - {\it
m}_{{\kern 1pt} \circ} v_{ \circ} ^{2} \cdot \left( {1{\kern 1pt}
{\kern 1pt} {\kern 1pt} -  \left( {v{\kern 1pt} /{\kern 1pt} v_{
\circ} } \right)^{2}} \right)^{1/2}.$

By analogy with special relativity, $m, {\kern 2pt} {\rm l}$ and
$\rm v$, are the parameters of a moving object, while $v_{ \circ}
\equiv c$ is the speed of light.

The existence of inertons was successfully verified experimentally (see,
e.g. Krasnoholovets, 2002, 2003; Bounias and Krasnoholovets, 2003b). Thus
inertons are basic excitations of the tessellattice, or in other words, they
are field particles, which provide for the quantum mechanical interaction,
the gravitational interaction and the inertial interaction (this one is
caused by so-called forces of inertia and the centrifugal force, i.e., in
this case inertons appear due to resistance to the motion of any object on
the side of the space).

\subsection*{7. Structural Classes of Particled Cells}

\subsubsection*{7.1. Particle-like Components}

\hspace*{\parindent} Denote by $\rm \theta = \{ \rho,{\kern
1pt}a,{\kern 1pt} I\} $ a quantum of fractality where $\rm I{\kern
1pt} {\kern 1pt} = \sum\nolimits_{{\kern 1pt} i\, = 1{\kern 1pt}
\to {\kern 1pt} {\kern 1pt} \infty}  {\{ 1/2^{i}\}} $ is the
initiator, $\rho $ is the self-similarity ratio and $a$ is the
additional number of subfigures inserted in the $\left( {1/\rho}
\right)$ fragments of the initial figure. The corresponding
fractal structure is denoted as $\left( {\Gamma} \right)$ that can
be decomposed in a sequence of elementary components $\rm \{ C_{1}
,C_{2} ,\quad ...,C_{k} ,...\} .$ If all these elementary
deformations are gathered on one single ball, then this ball
contains all the quantum of fractality, though its dimension is
not changed. It is therefore non-massive as it stands and its
motion is determined by the velocity of transfer of non-massive
deformations, that is the maximum permitted by the elasticity of
the space lattice. Since the deformations are ordered and
distributed in one particular structure, it owns a stability
through mappings of Poincar${\rm \acute{e}}$ sections. Such
particles are likely to correspond to bosons, i.e., to pseudo
particles representing transfer of packs of deformations in an
isolated form.

Hence, photon-like corpuscles will carry the equivalent of various
quanta of fractality $\rm \{ \vartheta_i\}_i,$ i.e., their
equivalent in mass in a decomposed form. This represents as many
deformations of the lattice, and finally of equivalent in energy.

\subsubsection*{7.2. Charges}

\hspace*{\parindent} These are opposite kinds of particle
deformations. When a quantum of fractal deformations collapses
into one single ball, two adjacent balls exhibit opposite forms:
one in the sense of convexity  and the other in the sense of
concavity on the surface of the ball. Hence, there occurs a pair
instead of a single object. The paired structures hold the same
fractal dimension, and they will retain the same masses if they
get the same volumes. This is realized if the member of the pair
whose deformation is in the convex sense looses an equivalent
volume in a nonfractal form.

The progression of such structures in the degenerate space will
generate several kinds of inerton cloud equivalent, depending on
convexity trends $\left( {\Xi}  \right)$ and symmetry properties
$\left( {\Psi}  \right)$ of the corresponding structures. The
properties generated by $\rm Q = \{ \Xi ,{\kern 1pt} {\kern 1pt}
\Psi \} $ can be called "charge effects" (regarding further
effects including the magnetic properties see Bounias and
Krasnoholovets, 2003a; Krasnoholovets, 2003).

\subsubsection*{7.3. Families of Massive Particles}

\hspace*{\parindent} Any single ball carrying a group $\{ \Gamma
{\kern 1pt} i\} {\kern 1pt} i$ of quantum fractals will represent
a class of massive particles. Depending on both the number and the
mode of association of these fractal quanta, various symmetries
will result and provide these classes with specific properties.

Simpler particles made from one single quantum of fractality $\rm
\{ \rho ,{\kern 1pt} a\} $ would likely correspond with
lepton-like structures, such that (here $N{\kern 1pt} i$ are odd
numbers) in{equation}
  $$
\rm  L_ {{\kern 1pt }i} = \left( \{ N_{\kern 1pt i} \cdot \left(
\rho_{\kern 1pt i} \right)^{e_i} \} \right).
  \eqno(28)
  $$

Hadron-like families will thus be represented by the following
common structures
  $$
\rm  H_{i,{\kern 1.5pt} k}  = ( \{ a_{\kern 1pt i}, {\kern 1pt}
\rho_{\kern 1pt i}, {\kern 1pt } e_{\kern 1pt i} \} _{i, {\kern
1pt} k} ).
  \eqno(29)
  $$

\subsubsection*{7.4. Spins for Balls}

\hspace*{\parindent} \textit{Fermion-like Cases}. Moving massive
balls have been shown to carry a cloud of deformations transferred
to degenerate balls of the surrounding space, with periodic
exchange between this inertons cloud and the original particle.
The spatial period of this pulse has been identified with the de
Broglie wavelength. Hence, the center of mass $\left( {{\kern 1pt}
y} \right){\kern 1pt} $ of the system composed of the particle and
the inerton cloud permanently undergoes a movement forwards and
backwards along the trajectory of the system. Two canonical
positions are possible, with respect to the particle: (i) $\left(
{y} \right){\kern 1pt} $ is centered on the particled ball, and
(ii) $y$ is no longer centered. The probability of state of is
thus ${\rm P}\left( {{\kern 1pt} y} \right){\kern 1pt} {\kern 1pt}
{\kern 1pt} = 1/2.$

\textit{Boson-like Cases}. Consider a ball carrying quanta of
masses in the decomposed form: then, such a system is opposed the
minimum resistance by the surrounding degenerate balls, which are
of the same nature, excepted that their individual densities of
deformation are much smaller. Therefore, boson-like particles do
not generate a cloud similar to that of a massive particle, and
their center of mass $\left( {y} \right){\kern 1pt} $ owns only
one main state: thus ${\rm P}\left( {y} \right) = 1.$

\textit{Spin Module}. The state of the center of mass is assessed
by the expected moment of junction $\rm \langle MJ \rangle $ of
its components, so that the spin-like system is described by ${\rm
P}\left( {x,y} \right){\kern 1pt} \langle m\rangle,$ standing for
$s \hbar /2,$ i.e., is the classical spin module expression (see
also Krasnoholovets, 2000).

\subsection*{8. Expansion of the Universe}

\hspace*{\parindent} In any one Poincar${\rm \acute{e}}$ section,
representing a timeless instantaneous state (an instant) of
universe, the lattice of space is represented by a stacking of
balls with nonidentical shape.

\textbf{Proposition.} Elementary balls exhibit increasing volumes from the
center to the periphery of a 3-D stacking.

Three arguments concur to the same proposition:

(i) oscillating deformations in excess in one cell can be partly
compensated by transfer to neighboring balls, like an equivalent
to the inerton cloud surrounding a particled ball. However, in
central parts, the volume available is limited by the density of
the stacking, and this limit is likely decreasing while going to
the outer coats of the lattice. Can this volume be defined
mathematically? This is a very specific question and the answer to
it should also take into account the shape and symmetry of cells
in the place of space studied.

In a simple estimation, we denote by (a) the radius of the
canonical (smallest) volume that can be transferred from a ball to
another. Assuming that each cell forwards a volume (a) to another
situated closer to the periphery, in the stacking, then the radius
of a ball in the n$^{\rm th}$ coat is approximated by:
  $$
\rm r_{\,n} = r_{1} + \left( {n{\kern 1pt} -  {\kern 1pt} 1}
\right) a; \eqno(30)
  $$

(ii) while the above considerations a valid for a particless lattice, if the
lattice is filled with particled balls, then there results a kind of
pressure due to the inerton clouds;

(iii) in contrast with the finiteness of volume to be compensatively
distributed in the surrounding cells, the area of a particled cell is
virtually infinite, and the needed area cannot be compensated by a finite
number of the surrounding cells. Thus an influence of any particle is likely
to be found up to the most remote parts of the lattice.

\subsection*{9. Conclusion}

\hspace*{\parindent} In this study, we have tried to pose as few
postulates as possible examining principles of constitution of
space. Generalized concepts of distances and dimensionality
evaluation are proposed, together with their conditions of
validity and range of application to topological space. The
mathematical lattice of empty sets (the tessellattice) provides a
substratum with both discrete and continuous properties, from
which the existence of a physical universe is derived. This is a
new theory of space whose physical predictions easily suppress the
opposition of both quantum and relativistic approaches, because
submicroscopic mechanics of particles derived from the theory
easily results in the Schr\"odinger's and Dirac's formalisms
(Krasnoholovets, 2000) and the gravitation phenomenon is deduced
from this submicroscopic mechanics as well, which will be shown in
further research.

Discrete properties of the lattice allow the prediction of scales
on which microscopic to cosmic structures should occur.
Deformations of primarily cells by exchange of empty set cells
allow a cell to be mapped into an image cell in the next section
as far as mapped cells remain homeomorphic. If a deformation
involves a fractal deformation to objects, the change in the
deformation of the cell takes place and the homeomorphism is not
conserved. The fractal kernel stands for a particle and the
reduction of its volume, which is associated with the appearance
of mass, is compensated by morphic changes of a finite number of
surrounding cells. Quanta of distances and quanta of fractality
have been demonstrated. It is shown that the interaction of a
particle-like deformation with the surrounding lattice results in
a fractal decomposition process, which strongly supports
previously postulated clouds of inertons as associated to moving
particles. Families of actual particles and field particles have
been analyzed.

The theoretical reasoning presented sheds light on the general
problem of hypothetical "origin" of the universe. The research
conducted brings out the sheer inconsistency of the Big Bang
concept that is still treated today as the only one possible for
cosmology. There was no any Big Bang in the remote past. The
universe is not an empty space or a vague vacuum, but it is a
substrate that eternally exists in the form of the tessellattice
and this is the simple truth, because the plain evidence of facts
is superior to all declarations.

The author (V.K.) is very thankful to the referee for the
constructive remarks that allowed improvements of the contents of
the paper possible.

\newpage

\subsection*{References}

\begin{description}
\item Borel, E., 1912. Les ensembles de mesure nulle. In: Oeuvres
completes. Editions du CNRS, Paris, Vol. 3, 1912.

\vspace{-4mm} \item Bounias, M., 2000. The theory of something: a
theorem supporting the conditions for existence of a physical
universe, from the empty set to the biological self. In: CASYS'99
In: CASYS'1999, International Conference on Computing Anticipatory
Systems (Daniel M. Dubois, ed.): Int. J. Comput. Anticipatory
Systems, 5, 11-24.

\vspace{-4mm} \item  Bounias, M., 2001. Undecidability and
Incompleteness In Formal Axiomatics as Questioned by Anticipatory
Processes. In: CASYS'2000, International Conference on Computing
Anticipatory Systems (Daniel M. Dubois, ed.): Int. J. Comput.
Anticipatory Systems, 8, 259-274.

\vspace{-4mm} \item Bounias, M., Bonaly, A., 1996. On metrics and
scaling: physical coordinates in topological spaces. Indian
Journal of Theoretical Physics, 44(4), 303-321.

\vspace{-4mm} \item Bounias, M., and Bonaly, A., 1997. Some
theorems on the empty set as necessary and sufficient for the
primary topological axioms of physical existence. Physics Essays,
10 (4), 633-643.

\vspace{-4mm} \item Bounias, M., and Krasnoholovets, V., 2003a.
Scanning the structure of ill-known spaces: Part 1. Founding
principles about mathematical constitution of space. Kybernetes:
The International Journal of Systems and Cybernetics, 32 (7/8),
945-975 (also http://arXiv.org/abs/physics/0211096).

\vspace{-4mm} \item  Bounias, M., and Krasnoholovets, V., 2003b.
Scanning the structure of ill-known spaces: Part 2. Principles of
construction of physical space, in Kybernetes: The International
Journal of Systems and Cybernetics, 32 (7/8), 976-1004 (also
http://arXiv.org/abs/physics/0212004).

\vspace{-4mm} \item  Bounias, M., and Krasnoholovets, V., 2003c.
Scanning the structure of ill-known spaces. Part 3. Distribution
of topological structures at elementary and cosmic scales, in
Kybernetes: The International Journal of Systems and Cybernetics,
32 (7/8), 1005-1020 (also http://arXiv.org/abs/physics/0301049).

\vspace{-4mm}  \item  Krasnoholovets, V., 1997. Motion of a
relativistic particle and the vacuum. Physics Essays 10(3),
407-416 (also http://arXiv.org/abs/quant-ph/9903077).

\vspace{-4mm}  \item  Krasnoholovets, V., 2000. On the nature of
spin, inertia and gravity of a moving canonical particle. Indian
Journal of Theoretical Physics 48 (2) 97-132 (also   \break
http://arXiv.org/abs/quant-ph/0103110).

\vspace{-4mm}  \item  Krasnoholovets, V., 2002. Submicroscopic
deterministic quantum mechanics, International Journal on
Computing Anticipatory Systems 11, 164-179 (CASYS'2001, Liege, ed.
D. Dubois) (also http://arXiv.org/abs/quant-ph/0109012).

\vspace{-4mm}  \item  Krasnoholovets, V., 2003 (or 2004). On the
nature of the electric charge. Hadronic Journal, in press.

\vspace{-4mm}  \item  Krasnoholovets, V., 2004. On the origin of
conceptual difficulties of quantum mechanics. In: Progress in
Quantum Physics Research, ed. V. Krasnoholovets. Nova Science
Publishers Inc., New York, 2004, to be published.

\vspace{-4mm}  \item  Marcer, P., Mitchell, E., and Schempp, W.
Self-reference, the dimensionality and scale of quantum mechanical
offers, critical phenomena, and qualia. International Journal on
Computing Anticipatory Systems (CASYS'2003, ed. D. Dubois), to be
published.

\vspace{-4mm}  \item  Nottale, L. 1997. Scale-relativity and
quantization of the universe I. Theoretical framework, {\it
Astron. Astrophys}. {\bf 327}, 867- 889.

\vspace{-4mm} \item  Nottale, L. 2003. Fractal space-time,
nondifferentiable geometry and scale relativity. Proc. Symposia in
Pure Mathematics, Special Volume "Fractal Geometry and
Applications: A Jubilee of Benoit Mandelbrot", in press.

\vspace{-4mm} \item  Tricot, C., 1999. Courbes et dimension
fractale. Springer-Verlag, Berlin Heidelberg, 240-260.
\end{description}

\end{document}